# Approximate Bayesian computation (ABC) coupled with Bayesian model averaging method for estimating mean and standard deviation

Deukwoo Kwon[1*]

*Corresponding author

Email: DKwon@med.miami.edu

Isildinha M. Reis[1,2]

Email: ireis@med.miami.edu

[1]Sylvester Comprehensive Cancer Center, University of Miami, Miami, FL 33136

[2]Department of Public Health Sciences, University of Miami, Miami, FL 33136

**Abstract**

**Background:** Previously (Kwon and Reis 2015), we proposed approximate Bayesian computation with single distribution selection (ABC-SD) for estimating mean and standard deviation from other reported summary statistics. We showed that ABC-SD outperforms other available methods when data are generated from skewed or heavy-tailed distributions and has comparable performance in symmetric distributions. The ABC-SD generates pseudo data from a single parametric distribution thought to be the true distribution of underlying study data. This single distribution is either an educated guess, or it is selected via model selection using posterior probability criterion for testing two or more candidate distributions. Further analysis indicated that when model selection is used, posterior model probabilities are sensitive to the prior distribution(s) for parameter(s) and dependable on the type of reported summary statistics.

**Method:** We propose ABC with Bayesian model averaging (ABC-BMA) methodology to estimate mean and standard deviation based on various sets of other summary statistics reported in published studies. We conduct a Monte Carlo simulation study to compare the new proposed ABC-BMA method with our previous ABC-SD method.

**Results:** In the estimation of standard deviation, ABC-BMA has smaller **average relative errors (AREs)** than that of ABC-SD for normal, lognormal, beta, and exponential distributions. For Weibull distribution, ARE of ABC-BMA is larger than that of ABC-SD but <0.05 in small sample sizes and moves toward zero as sample size increases. When underlying distribution is highly skewed and available summary statistics are only quartiles and sample size, ABC-BMA is recommended but it should be used with caution. Comparison of mean estimation between ABC-BMA and ABC-SD shows similar patterns of results as for standard deviation estimation.

**Conclusion:** ABC-BMA is easy to implement and it performs even better than our previous ABC-SD method for estimation of mean and standard deviation.

## Background

Hozo et al. [1], Bland [2], and Wan et al. [3] proposed algebraic methods to estimate the sample mean and standard deviation. For instance, Hozo et al. [1] proposed a simple method for estimating the sample mean and standard deviation from the median, minimum, maximum, and the sample size. Bland [2] proposed estimating these quantities based on the three quartiles, minimum, maximum, and the sample size. More recently, Wan et al. [3] proposed improved versions for Hozo et al. and Bland methods. They also provided a new method for estimation of the sample standard deviation based on the three quartiles, and the sample size.

As an alternative to the above methods, we recently proposed a Monte Carlo procedure, the approximate Bayesian computation with single distribution selection (ABC-SD) method (Kwon and Reis [4]), for estimating mean and standard deviation. In our previous simulations studies [4], we showed that ABC-SD outperforms other available methods when data are generated from skewed or heavy-tailed distributions and has comparable performance in symmetric distributions. We showed that the ABC-SD method produces more precise estimates of true study-specific mean and standard deviation as sample size increases and it also accommodates various distributions for underlying original data. Furthermore, application of ABC-SD is not restricted to having available the mentioned descriptive statistics (minimum, maximum, and/or quartiles). ABC-SD can be applied using other reported summary statistics such as mean and asymmetric 95% confidence interval, or Bayesian summaries (e.g.: posterior mean, the maximum a posteriori probability (MAP), and 95% credible interval).

The ABC-SD method generates pseudo data from a single distribution thought to be the distribution of unavailable study data. This single distribution is either an educated guess or it is selected via model selection using posterior probability criterion for testing two or more candidate distributions. When model selection is used, we found out that posterior model probabilities are sensitive to the prior distribution(s) for parameter(s) and dependable on the type of reported summary statistics. In certain situations, model selection may cause misspecification of underlying distribution for ABC-based estimation method and lead to biased estimates of mean and standard deviation.  As a solution for these early observations, we now proposed ABC with Bayesian model averaging (ABC-BMA) to estimate sample mean and standard deviation. We found that our new proposed ABC-BMA method gives quite small relative errors compared to that of ABC-SD. An example of this is illustrated in section "***Sensitivity analysis of prior distribution setting***" where we generate data from beta, and examine model selection between beta and normal as the potential underlying distributions of sample data, using different prior distribution settings for parameters.

In 'Methods' section we describe the proposed ABC-BMA method and its computational implementation. We begin with a brief summary of the ABC-SD method (Kwon and Reis [4]). In 'Results', we describe and report the findings of the simulation studies comparing the performance of these methods and the impact of prior distributions.  We used the statistical software R in performing all statistical programming related to the implementation of the various methods, analysis, and simulations.

## Methods

Meta-analysis is usually based on summary (aggregate) data from selected publications, in the form of a handful of descriptive or other summary statistics reported. Thus, in this context, formal Bayesian inference cannot be implemented due to unavailable of the subjects' original research data required for estimation of the data likelihood. However, using ABC methods, the calculation of the likelihood can be replaced by a comparison of summary statistics from the observed data and those from simulated pseudo data using a distance measure. Several review papers provide more detailed descriptions of ABC method as indicated in Beaumont [5] and Marin et al. [6].

**ABC-SD: ABC with single distribution selection**

Here we give a brief summary of our ABC-SD method (Kwon and Reis [4]) for estimation of the mean and standard deviation using summary statistics. As shown in Table 1 (left column) ABC-SD begins by choosing a single parametric distribution as the potential underlying distribution of original study data, $f(\theta)$. We can choose an underlying distribution based on educated guess, given the set of reported summary statistics and the nature of outcome variable. For instance, when the outcome is related to distribution with positive support, there are several distributions to be considered, such as log-normal, Weibull, or exponential.

After an underlying distribution selection, the second step of ABC-SD is to generate candidate value(s) for parameter(s), $\theta^*$, from uniform prior distribution(s), $p(\theta)$. For example, if we choose as the potential underlying distribution of original data the exponential distribution, then we only need to specify one uniform prior distribution for the scale parameter, $\lambda$. If we choose a normal distribution, then two independent prior uniform distributions are need for the corresponding parameters mean ($\mu$) and standard deviation ($\sigma$). The third step is to generate pseudo-data, $D^*$, from the chosen underlying distribution with likelihood function $f(\theta^*)$, and to compute the same reported summary statistics of the observed data, $S(D)$, from pseudo data, $S(D^*)$. The fourth and last step is to decide whether θ* is accepted or not. This decision depends on the distance between summary statistics of the observed data, S(D), and those of pseudo data, S(D*) denoted by ρ(S(D),S(D*)), where ρ(•,•) is a distance measure. In our application of ABC, we used the Euclidean distance measure. If ρ(S(D),S(D*)) is smaller than a fixed tolerance value ε (i.e., ρ(S(D),S(D*))<ε), then θ* is accepted, otherwise it is rejected. Steps 2-4 are repeated a large number of times (e.g., N=20,000) in order to obtain a big set of θ* values for the inference. Alternatively, instead of setting a small tolerance value, ε, we can select a fixed number of θ* values corresponding to an acceptance percentage. For example, with acceptance percentage of 0.1% and N=50,000, we select 50 values of θ* corresponding to the top 0.1% with smallest Euclidean distance. The fundamental idea of ABC is that a good approximation of the posterior distribution can be obtained using summary statistics, S(D), and a fixed small tolerance value ε (or a pre-specified acceptance percentage). Mean and standard deviation are estimated using accepted values of parameter(s) in ABC-SD. In our previous publication (Kwon and Reis [4]) we proposed two ways to estimate mean and standard deviation: 'plug-in method' and 'simulation'. The plug-in method consists of replacing means of accepted parameter values into the corresponding formulas for the mean and standard deviation. The simulation approach consists of obtaining the mean values of calculated mean and standard deviation during steps 2-4 implementation of ABC-SD. These two methods provide similar estimates of mean and standard deviation. In some applications/situations it may be difficult to choose a unique underlying distribution. In this case, as described in our previous publication, we implement ABC-SD method for the several candidate distributions (K>2). Employing model selection (i.e. distribution selection in our context) while we apply the ABC-SD method, we calculate corresponding marginal posterior model probabilities (P(M_k|S), k=1,...,K) and then choose the

distribution with the highest marginal posterior model probability among K candidate distributions.

## ABC-BMA: ABC coupled with Bayesian model averaging method

Although ABC-SD outperformed the previous proposed methods by Hozo et al. [1], Bland [2], and Wan et al. [3], we found that the selection of underlying distribution via posterior model probability was sensitive to the prior distribution(s) for parameter(s) and available descriptive summary statistics. To improve estimation, we now propose the use of ABC with Bayesian model averaging (ABC-BMA) methodology, using posterior model probability as weights, to estimate mean and standard deviation based on various sets of summary statistics reported in published studies. Briefly, under ABC-BMA, estimates of mean and standard deviation are weighted average of those accepted values (estimated means and standard deviations of pseudo data). The weights are the proportions of accepted values from each candidate distribution, which are the estimated posterior model probabilities for the distributions.

Table 1 (right column) summarizes the steps for implementation of the ABC-BMA method. The first step is to produce a list of K potential candidate distributions (K>1). The second step of ABC-BMA is to choose a distribution among the K distributions ($M_1$, ..., $M_K$) using multinomial distribution with probability $\underline{\pi} = (\pi_1, ..., \pi_K)$, where $\pi_k=1/K$, k=1,...,K. In the third step, given chosen distribution $M_k$, we generate a set of candidate value(s) for parameter(s), $\theta^*$, from specific uniform prior distribution(s), $p_{M_k}(\theta)$. The fourth step is to generate pseudo data, D*, from the likelihood function, $f_{M_k}(\theta^*)$, where $f_{M_k}(\cdot)$ is a chosen underlying distribution and compute summary statistics of simulated data, S(D*). The fifth step is to decide whether $\theta^*$ is accepted or not. This decision is similar to that of ABC-SD, which depends on the distance between summary statistics of the observed data, S(D), and those of simulated data, S(D*) denoted by $\rho(S(D),S(D^*))$. Steps 2-5 are repeated a large number of times (e.g., N=20,000) in order to obtain estimates of mean and standard deviation and $M_k$ for the inference. In order to obtain better convergence we consider adaptive step during the iterations. The $\pi_k$ values in $\underline{\pi}$ will be adjusted at every 1,000 iterations according to accepted frequency of each distribution, $M_k$. As the iteration increases, a more promising distribution is likely to be selected more often. In the sixth and last step, we calculate mean and standard deviation estimates as the weighted average of those accepted values (estimated means and standard deviations of pseudo-data). As mentioned earlier, the weights are the proportions of accepted values from each candidate distribution.

## Results

### *Designs of simulation studies*

Previously in [4], we showed that ABC-SD method outperformed the other available methods (Hozo et al., Bland, and Wan et al.).The main focus in this simulation study is then to compare ABC-BMA with ABC-SD only. In this paper, we use the same simulation design settings used in [4], including the same five underlying data distributions (normal, log-normal, beta, exponential and Weibull) and the same three scenarios (S1, S2, and S3) of available summary statistics. That is, we assume available in S1, the minimum ($x_{min}$), median ($x_{Q2}$), maximum ($x_{max}$) and the sample size (n); in S2, the first and third quartiles ($x_{Q1}$ and $x_{Q3}$) in addition to summary statistics in S1; and in S3, the three quartiles and the sample size.

**In S1**, we use five distributions: normal distribution with mean 50 and standard deviation 17, N(50,17); log-normal distribution with location parameter=4 and scale parameter=0.3, LN(4,0.3); Weibull distribution with shape parameter=2 and scale parameter=35, Weibull(2,35), beta distribution with shape parameters 9 and 4, Beta(9,4); and exponential distribution with mean=10, Exponential(10). **In S2**, we compare two ABC-based methods in lognormal and beta distributions. We use three log-normal distributions with same location parameter value of 5 but having three different scale parameter values (0.25, 0.5, and 1) in order to evaluate effect of skewness. In addition, we use three beta distributions, Beta(5,2), Beta(1,3), and Beta(0.5,0.5), to examine effect of skewness and bimodality in estimation for bounded data distribution. **In S3**, we use the same five distributions in S1 (normal, lognormal, beta, exponential and Weibull) to investigate further the effect of chosen descriptive statistics for the standard deviation estimation. In each scenario we consider 10 sample sizes (n= 10, 40, 80, 100, 150, 200, 300, 400, 500, 600). We obtain a sample of n observations from a particular distribution, and compute the sample mean (true $\bar{x}$) and sample standard deviation (true S), as well as sample descriptive statistics under each scenario. We obtain the estimates of the sample mean and standard deviation from the given sample descriptive statistics under each scenario, using the two ABC methods. The relative errors (REs) are calculated as follows:

$$RE \ of \ mean = \frac{(estimated \ \bar{x} - true \ \bar{x})}{true \ \bar{x}}, \tag{1}$$

and

$$RE \ of \ standard \ deviation = \frac{(estimated \ S - true \ S)}{true \ S}. \tag{2}$$

For each sample size n, we repeat this procedure 200 times to obtain average relative errors (AREs).

In the simulations, we set acceptance percentage 0.1% and 20,000 total number of iterations for ABC-SD method and same acceptance percentage of 0.1% and 50,000 total number of iterations for ABC-BMA. For fair comparison, we use the same uniform prior distributions for each distribution as described in Table 2 of Kwon and Reis [4]. For ABC-BMA, we set as our list of potential candidate underlying distributions the following: normal, lognormal, exponential, Weibull, and beta distributions.

### Results of simulation studies

In the simulation studies we compare estimation performance of the two ABC-based methods in terms of average relative error (ARE) for estimating mean and standard deviation. In the next three subsections we present comparison of methods for standard deviation estimation. In the last subsection, we present comparison of the two ABC-based methods for mean estimation.

### Comparison of ABC-SD and ABC-BMA for standard deviation estimation in scenario S1

Figure 1 shows that AREs in estimating standard deviation for ABC-SD and ABC-BMA methods for five different distributions. When the underlying distribution is normal, lognormal, beta, or exponential, ABC-BMA has much smaller AREs in small sample size (<100) and comparable AREs in all other sample sizes. For lognormal and beta distributions, it seems to be huge improvement in reducing AREs in small and moderate sample sizes when we use the ABC-BMA method. For Weibull distribution, ARE of ABC-BMA is larger than that of ABC-SD but <0.05 in small sample sizes and moves toward zero as sample size increases. Overall, for non-symmetrical distributions, ABC-BMA method shows improvement in reducing AREs in small and moderate sample sizes and maintain small AREs in large sample sizes.

***Comparison of ABC-SD and ABC-BMA for standard deviation estimation in scenario S2***

In this scenario, S2, more information is available compared to other two scenarios; that is, we assume available the minimum ($x_{min}$), maximum ($x_{max}$), the three quartiles ($x_{Q1}$, $x_{Q2}$ and $x_{Q3}$) and sample size (n). We examine performance of two ABC-based methods for asymmetric distributions, either highly skewed or bimodal distribution. Using lognormal distribution, we investigate change of AREs as degree of skewness increases from mild to severe. The ABC-SD method showed decreasing ARE pattern, although relatively large AREs were shown in small and moderate sample sizes. The ABC-BMA method performs better than ABC-SD in small or moderate sample sizes. The two methods are similar in large sample size, but ARE of ABC-BMA is more close to 0. (Figure 2).

Using beta distribution, we investigate the effect of bimodality as well as skewness for bounded data. (Figure 2) The two methods show negative AREs, that is, underestimation of study-specific true sample standard deviation. For all three beta distributions, the two methods have quite similar AREs when sample size is greater than 40. The ABC-BMA performs better and shows gain in terms of ARE towards zero for n≤40.

These results indicate that the ABC-BMA method performs better than ABC-SD in asymmetric distributions, and also better than the method by Wan et al. [3], shown in our previous publication [4], in asymmetric distributions when sample size is small and methods are similar in large sample sizes.

***Comparison of ABC-SD and ABC-BMA for standard deviation estimation in scenarios S1, S2, and S3***

We investigate the effect of having available various sets of summary statistics on the performance of the ABC-BMA method (Figure 3). For normal and beta distribution in S3, ABC-BMA shows very similar ARE pattern compared to that of ABC-SD. However, compared with other two scenarios (S1 and S2), ABC-BMA method in S3 has worse performance for small n, and similar performance for moderate to large n. For underlying lognormal distribution, when n>40 ARE is negative (that is, underestimation) for ABC-BMA in S3, in contrast to ABC-SD in S3 and the other ABC-BMA in S1 and S2 showing slight overestimation. Moreover, the ABC-BMA in S3 shows overall larger AREs as compared to ABC-SD under Weibull, and in small sample sizes under exponential. It seems that the $Q_1$, and $Q_3$ summary statistics in S3 are not sufficient to distinguish potential distributions with small to moderate skewness. For skewed distributions, having $x_{min}$ and $x_{max}$ improve estimation as shown by ARE close to zero in S1 and S2.

***Comparison of ABC-SD and ABC-BMA for mean estimation in scenarios S1, S2, and S3***

We compare AREs for mean estimation between the two ABC-based methods. In Figure 4, solid line denotes ABC-SD and dashed line denotes ABC-BMA. Circle symbols are for S1 (solid circle for ABC-SD and open circle for ABC-BMA), triangle symbols for S2 (solid triangle for ABC-SD and open triangle for ABC-BMA), and cross symbol is for S3. The two ABC-based methods have comparable performance in estimating mean for normal and beta distributions in all three scenarios. For exponential and Weibull distribution, ABC-BMA performs better than ABC-SD in small sample sizes under S1 and S2. While AREs of ABC-SD for lognormal distribution under S1 show large deviation from zero in moderate sample sizes, those of ABC-BMA are close to zero and show huge improvement. Like in standard deviation estimation, for moderately skewed distributions under S3 (log-normal, and Weibull for example), ABC-BMA shows slightly worse performance than ABC-SD in estimating mean.

### *Model probabilities of distributions in ABC-BMA method by available summary statistics*

We examine model probabilities of distribution selection across different sample sizes to elucidate effect of available summary statistics. Normal, beta, and exponential distributions have large average model probabilities (>0.8) in all scenarios. (See Figures 5A, 5B, and 5C). However, lognormal and Weibull distributions show different pattern of average model probability. For lognormal distribution, when sample size are less than 200, average model probability of lognormal distribution is less than 0.5 and that of normal distribution is larger than 0.5 in both S1 and S2. For Weibull distribution, normal and exponential distributions are main contender when sample size is small and average model probability of Weibull becomes larger than 0.5 when sample size is more than 40 under S1 and S2. These two distributions do not have meaningful average model probability for all sample sizes under S3. Instead, normal distribution has high average model probability (see Figure 5C).

When underlying true distribution is symmetric, highly skewed, or bounded support (e.g., normal, exponential, and beta distribution in our simulation study), true underlying distribution has very high model probability in all sample sizes. These shapes of underlying distributions can be easily identified by available summary statistics in all three scenarios. However, if a distribution is moderately skewed, like lognormal and Weibull distributions in our simulation setting, having only the 3 quartiles and sample size is not sufficient to distinguish from normal distribution. When $x_{\min}$ and $x_{\max}$ are available in addition to $Q_2$ or all 3 quartiles, moderately skewed underlying true distribution has high average model probability as sample size increases.

### *Sensitivity analysis of prior distribution setting*

We conduct a small simulation study for the sensitivity analysis of prior distribution setting. We consider underlying distribution is a beta distribution with shape parameters 9 and 4 and generate data for a sample size of 400 and calculate the following set of summary statistics: $x_{Q1}$=0.5993, $x_{Q2}$=0.6853, $x_{Q3}$=0.7735 sample mean=0.6814, and sample SD=0.1247. We examine model selection between beta and normal distributions with different prior settings and assuming available only the 3 quartiles and sample size ($S_3$). For prior distributions of the shape parameters in beta distribution we use same uniform distributions as follows: either U(0,40), used in the previous publication and current simulations, or U(0,20), which characterizes a less vague prior distribution. For normal distribution, we use for location parameter, μ, a fixed prior U($x_{Q1}$, $x_{Q3}$) and for scale parameter, σ, either U(0,1) or U(0,0.5). Note that the latter one represents less vague prior distribution. Based on these prior distributions for beta shape parameters and normal scale parameter, we have the following four combinations of priors for this simulation study: (1) U(0,40) for beta distribution and U(0,1) for normal distribution, (2) U(0,40) for beta distribution and U(0,0.5) for normal distribution, (3) U(0,20) for beta distribution and U(0,1) for normal distribution, and (4) U(0,20) for beta distribution and U(0,0.5) for normal distribution. In Table 2 we report posterior model probability for each distribution, along with relative errors (REs) for mean and SD from ABC-SD and ABC-BMA. In three out of four cases, beta distribution has model probability greater than 0.5. However, if normal distribution has less vague prior distribution for σ then model probability of normal is larger than 0.5, and it would be chosen as the underlying distribution of original sample data under ABC-SD. This would give more biased estimate of SD but not of mean. Moreover, it is not easy to assign similar vagueness for parameters of each distribution since some parameters are scale parameter and others are shape parameters. If we unintentionally assign less vague prior distribution of one of parameters, model probability might lead to incorrect choice of underlying distribution.

## Real Data Examples

We illustrate estimation of sample means and standard deviation, via our two ABC methods and also by the method of Wan et al. [3], using real data from two publications. For the ABC methods we use 0.1% acceptance percentage and total number of iterations is 100,000. In addition, for ABC-BMA we set five underlying distributions: normal, log-normal, beta, exponential and Weibull.

As our first illustration, we use data from a randomized controlled trial by Silver et al. [7], to estimate mean and standard deviation of the length of stay (LOS) in hospital. The available summary statistics for LOS in the 3% hypertonic saline arm, shown in Table 3 of [7], were sample size (n=111), median, $Q_2$ (2.1), the first quartile, $Q_1$ (1.2), and the third quartile, $Q_3$ (4.6) (S3 scenario). Wan et al. method, which assumes normal distribution, gives 2.633 for mean estimate and 2.554 for SD estimate. ABC-SD based on normal distribution gives 2.680 for mean estimate and 2.612 for SD estimate. These two methods give quite comparable estimates for both mean and SD, since both assume underlying data is normally distributed. ABC-SD based on log-normal gives 3.918 for mean estimate and 5.384 for SD estimate. Based on summary statistics, we can conjecture that underlying distribution is positively skewed (long right tailed), therefore estimates from ABC-SD based on log-normal are better than those from ABC-SD based on normal or Wan et al. method. ABC-BMA gives 3.802 for mean estimate and 5.205 for SD estimate with slightly higher model probability for lognormal distribution (52%). Estimates from ABC-SD based on log-normal and from ABC-BMA are similar, and both are more appropriated. However, as shown in the simulation ABC-BMA provides relative small bias in small sample sizes even if true underlying distribution is log normal.

In our second illustration, we use data by Hamstra et al. [8] on the Expanded Prostate Cancer Index Composite (EPIC), which is one of widely used patient-reported outcomes in prostate cancer population. EPIC consists of four domains: Urinary, Bowel, Sexual, and Hormonal domains. In Hamstra et al. [8], EPIC bowel summary scores at baseline, 2 months, 6 months, 12 months, and 24 months were reported with sample size, median, mean, minimum, and maximum values (S1 scenario). In this illustration, we estimate standard deviation for EPIC bowel summary score at the baseline. Reported summary statistics, in Table 2 of [8], for the baseline are 100 (median), 95 (mean), 91.7 (minimum), and 100 (maximum). Since EPIC score is bounded between zero and one hundred, we perform ABC-SD based on beta distribution with transformed scale falling into between zero and one. In addition, we perform ABC-SD based on normal distribution to be able to compare with Wan et al. method. The estimate of SD from ABC-SD based on beta distribution is 1.168, based on normal distribution is 1.528 and by Wan et al. method is 1.453. From ABC-BMA, estimate of SD is 1.233 and it is preferable based on results of our simulation studies on underlying beta distribution.

# Discussion

In general, ABC-based method to estimate mean and standard deviation is very flexible to the available summary statistics and performs well in various distributions. We rely on model probability of distribution selection to choose underlying distribution in ABC-SD. The model probability is sensitive to prior distribution setting and available summary statistics. The misspecification of underlying distribution for ABC-SD can lead to more biased (under or over)estimation of mean and standard deviation estimate. We show this in the particular simulation study comparing beta and normal distribution. As solution to this, we investigated and now proposed ABC-BMA which implements estimation as a weighted average of estimates from K>2 potential distributions for underlying data, using the corresponding posterior model probabilities as weights. We found that ABC-BMA gives quite small RE compared to that of single distribution, that is, our ABC-SD method

In order to understand why estimate from model averaging approach has smaller error than estimate from single distribution, we run a particular simulation study, setting n=100 and lognormal distribution with μ=4 and σ=0.3 in S1. Figure 6 shows individual relative errors (REs) for 200 repetitions of simulated data. ARE of ABC-SD is -0.186 and of ABC-BMA is 0.031, corresponding to 83.3% reduction in absolute bias. When data has an underlying lognormal distribution, normal distribution is a main contender for selection based on model probability in ABC-BMA (see Figure 5A, log-normal plot). In small sample sizes, we observe that model probability of normal distribution can be higher than that of lognormal distribution even though underlying true distribution is lognormal. In Figure 6 note that 90% of estimates of standard deviation from ABC-SD with lognormal distribution are under estimated. In ABC-BMA estimation, significant portion of estimates comes from normal distribution and those tend to be over-estimated and compensate under-estimated estimates from lognormal distribution.

## Conclusion

We propose an enhanced ABC-based approach utilizing BMA to estimate the mean and standard deviation when only descriptive statistics are available. Our ABC-BMA method shows reduction of AREs in small and moderate sample sizes and comparable performance to ABC-SD approach in large sample size. AREs of ABC-BMA method were smaller and very stable across different sample sizes when underlying distribution is less skewed. When underlying distribution is highly skewed and available summary statistics are $Q_1$, $Q_2$, and $Q_3$, ABC-BMA is recommended but it should be used with cautious. Comparison of mean estimation between ABC-BMA and ABC-SD showed similar patterns of results as for standard deviation estimation.

## Competing interests

The authors declare that they have no competing interests.

## Authors' contribution

DK and IR conceived and designed the methods. DK conducted the simulations. All authors were involved in the manuscript preparation. All authors read and approved the final manuscript.

## Acknowledgements

The authors would like to thank Dr. Robert C. Duncan for his helpful comments and proof-reading.

List of Figures

**Figure 1**.

Average relative error (ARE) comparison of ABC-SD and ABC-BMA methods in estimating sample standard deviation under S1. **A, B, C, D, E:** AREs for the two methods using simulated data from five distributions: normal, lognormal, beta, exponential, and Weibull, respectively. ABC-SD (solid circle with solid line) and ABC-BMA (open circle with dashed line).

**Figure 2**.

Average relative error (ARE) comparison of ABC-SD and ABC-BMA methods in estimating sample standard deviation under S2. **A, B, C:** AREs for the two methods using simulated data from three beta distributions. **D, E, F:** AREs for two methods using simulated data from three lognormal distributions. ABC-SD (solid circle with solid line) and ABC-BMA (open circle with dashed line).

**Figure 3**.

Average relative error (ARE) comparison of ABC-SD and ABC-BMA methods in estimating sample standard deviation under S1, S2, and S3. **A, B, C, D, E:** AREs for two methods using simulated data from five distributions: normal, lognormal, beta, exponential, and Weibull, respectively. ABC-SD method in S3 (black cross with solid line), ABC-BMA method in S1 (red open circle with dashed line), in S2 (red open triangle with dashed line), and in S3 (red cross with dashed line).

**Figure 4**.

Average relative error (ARE) comparison of ABC-SD and ABC-BMA in estimating sample mean under S1, S2, and S3. **A, B, C, D, E:** Simulated data from normal, lognormal, beta, exponential, and Weibull distributions**,** respectively. ABC-SD method in S1 (black solid circle with solid line), in S2 (black solid triangle with solid line), and in S3 (black cross with solid line). ABC-BMA method in S1 (red open circle with dashed line), in S2 (red open triangle with dashed line), and in S3 (red cross with dashed line).

**Figure 5**.

Average model probability in ABC-BMA method as function of sample size and underlying true distribution under S1 (A), S2 (B), and S3 (C). In each scenario, simulated data from five distributions: normal, log-normal, beta, exponential, and Weibull.

**Figure 6**.

Relative errors (REs) comparison in estimating sample standard deviation under S1using 200 simulated data sets of sample size 100 from lognormal distribution, LN(4, 0.3). ABC-SD method (black open circle) and ABC-BMA method (red solid circle). Two horizontal lines represent ARE for each method (black dashed line for ABC-SD and red dashed line for ABC-BMA).

List of Tables

**Table 1.** Procedures for implementation of ABC-SD and ABC-BMA methods

**Table 2.** Sensitivity of prior distributions on estimation of mean and standard deviation.

Figure 1

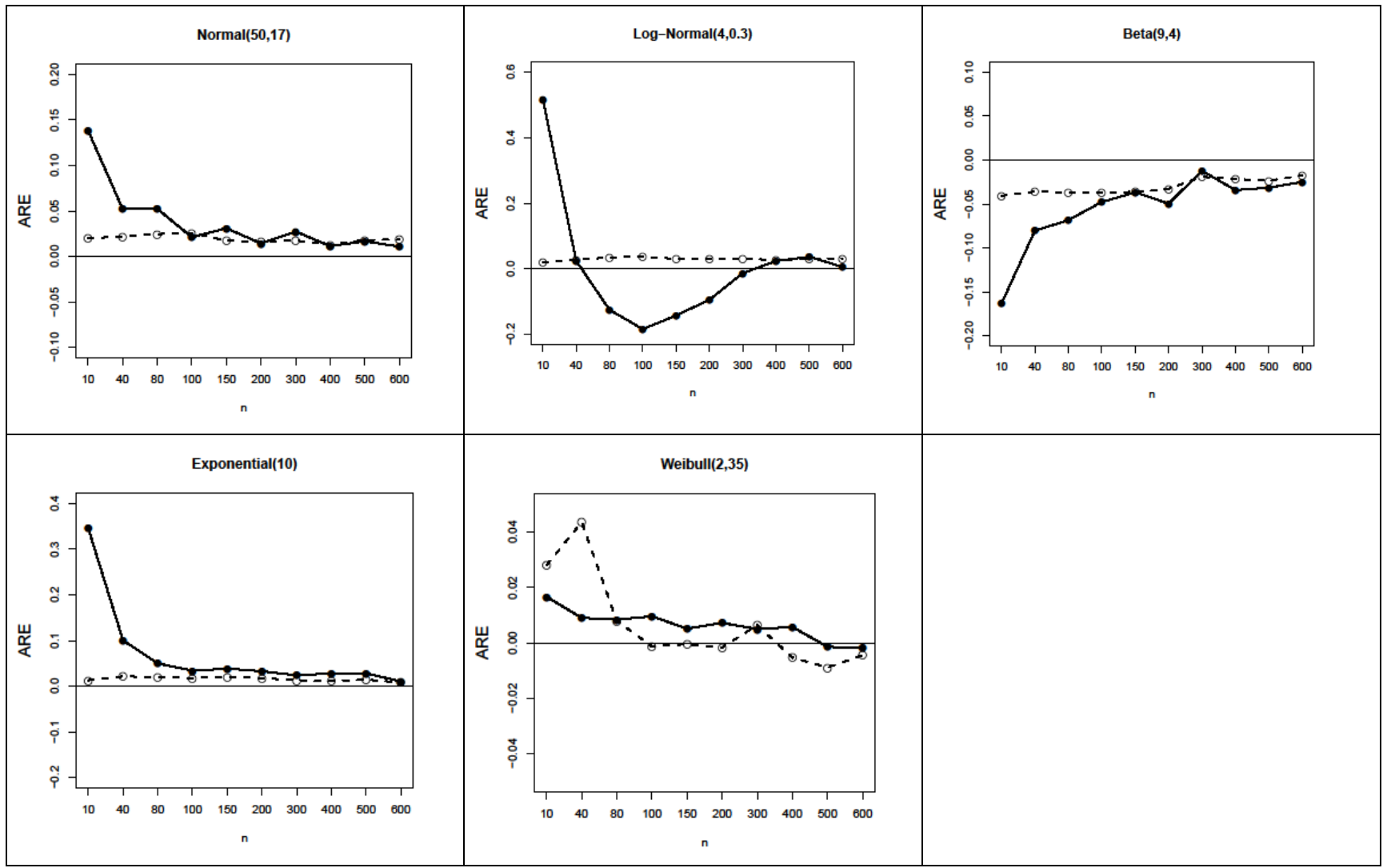

Figure 2

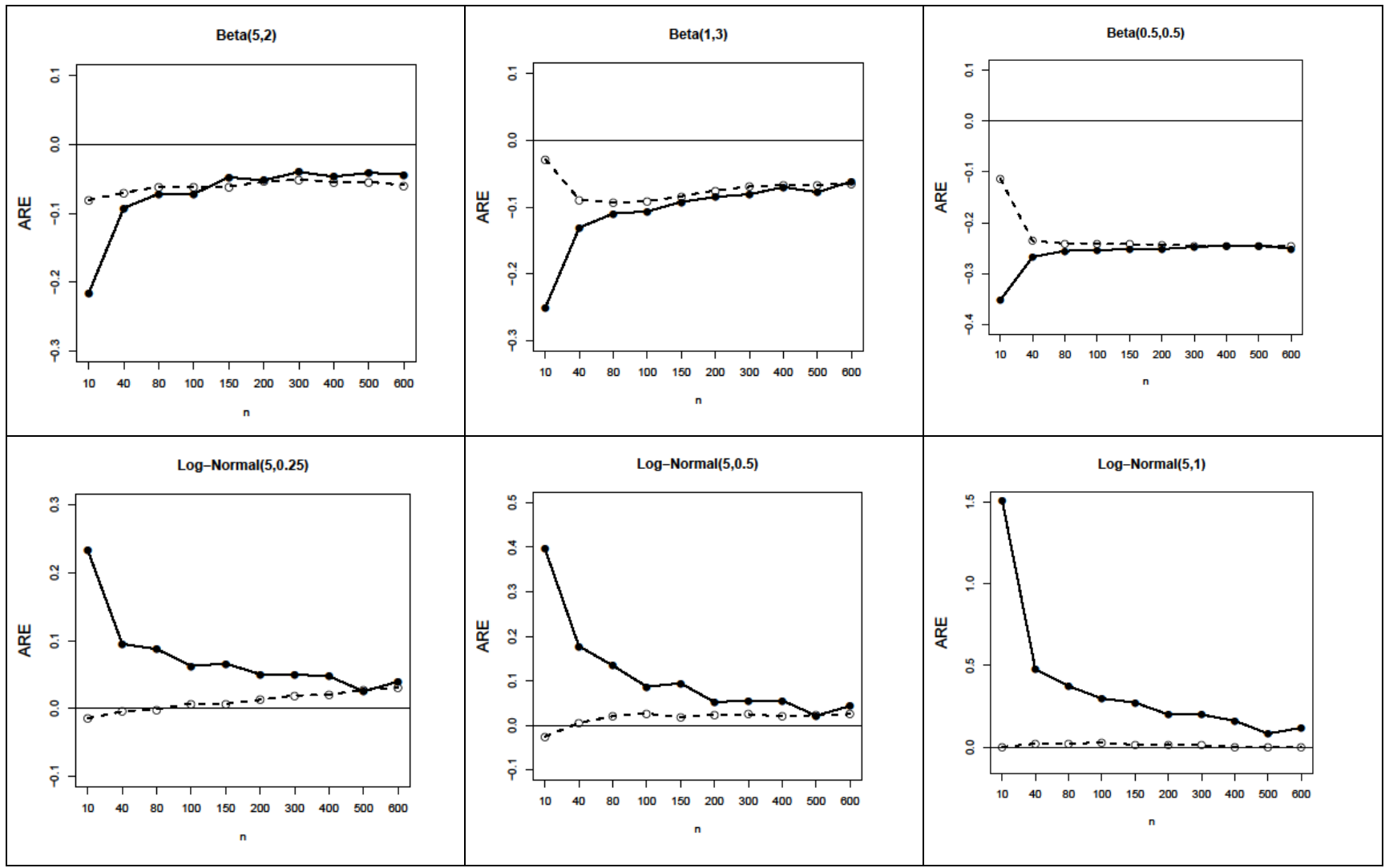

Figure 3

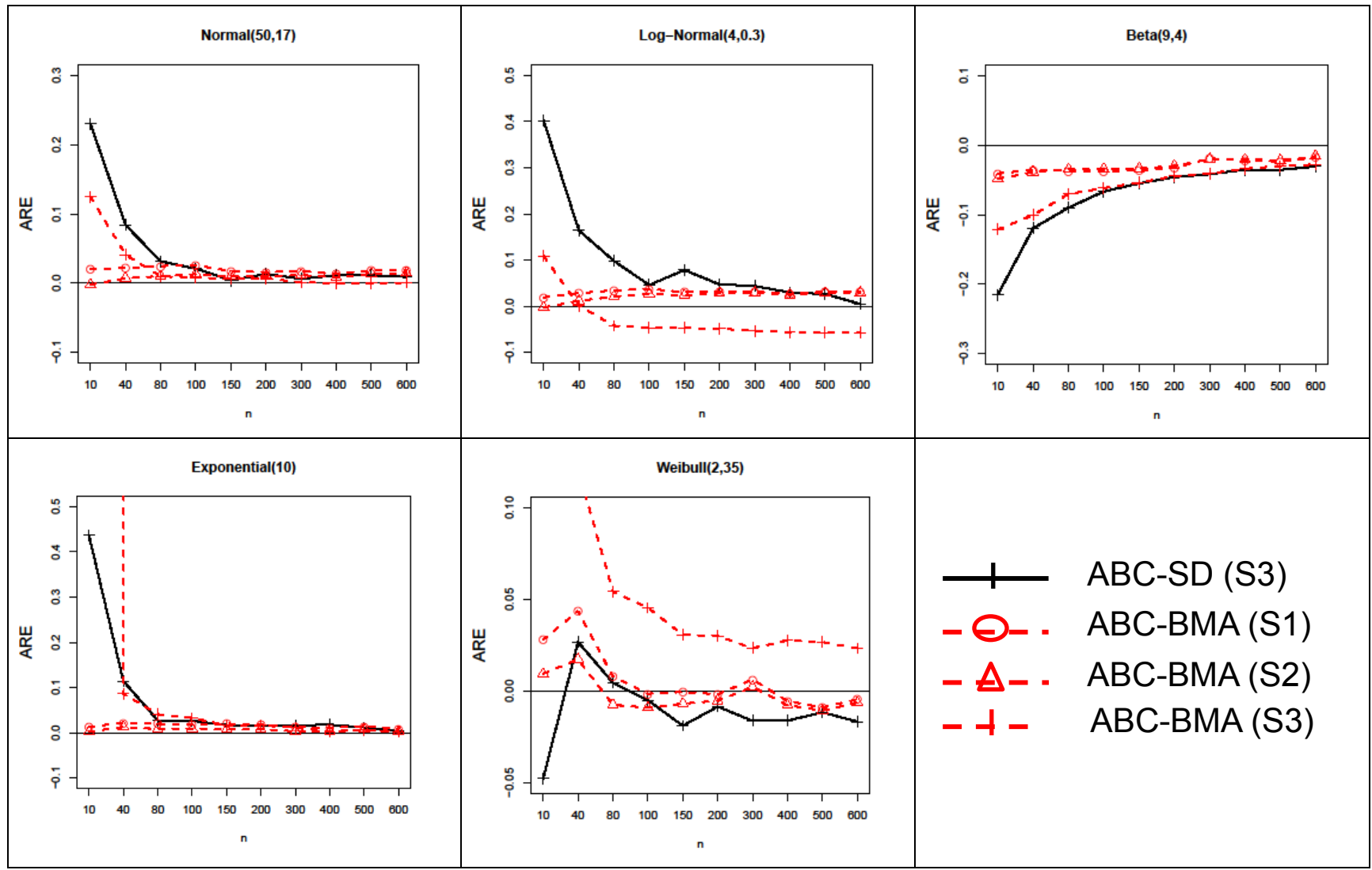

Figure 4

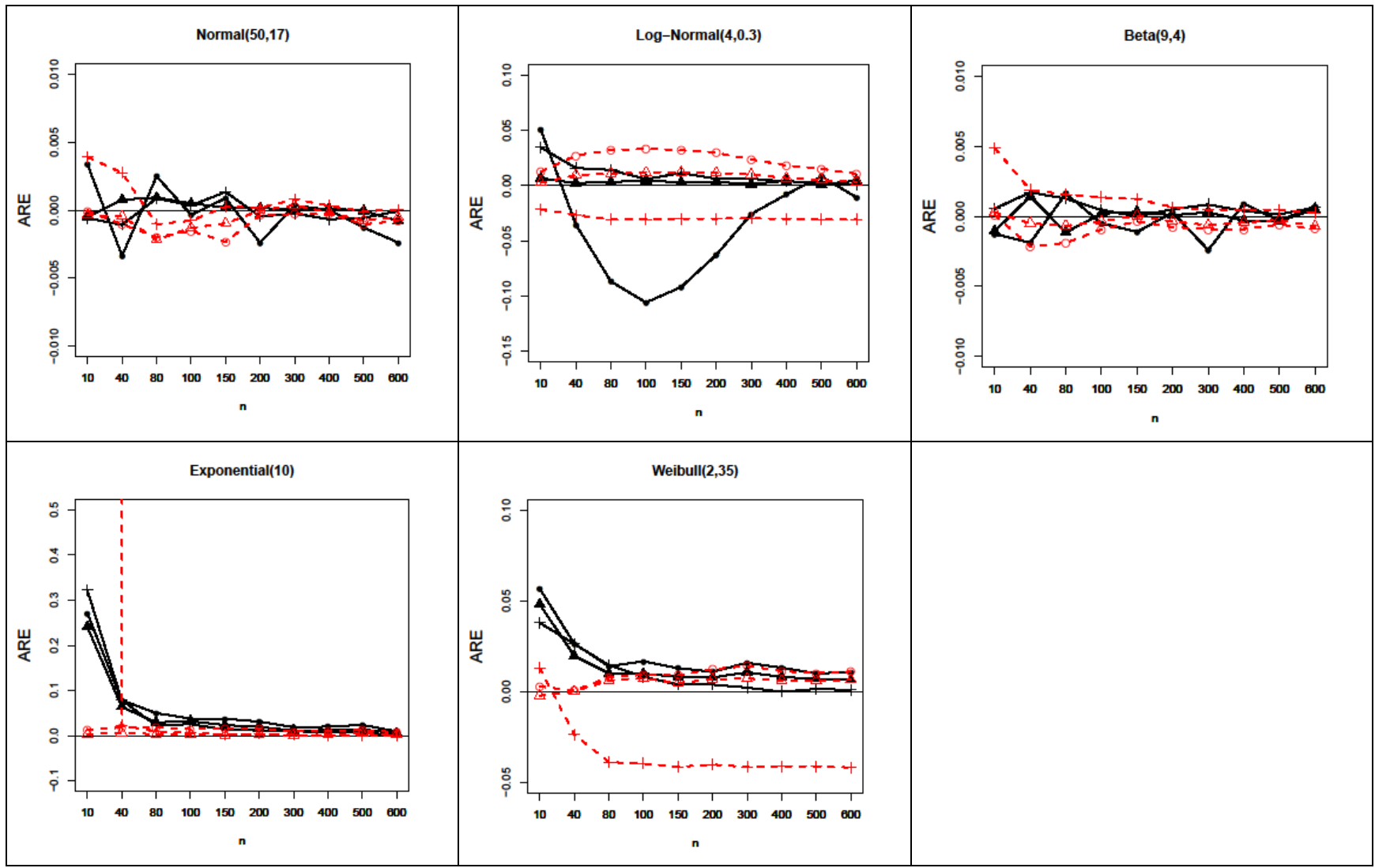

Figure 5A

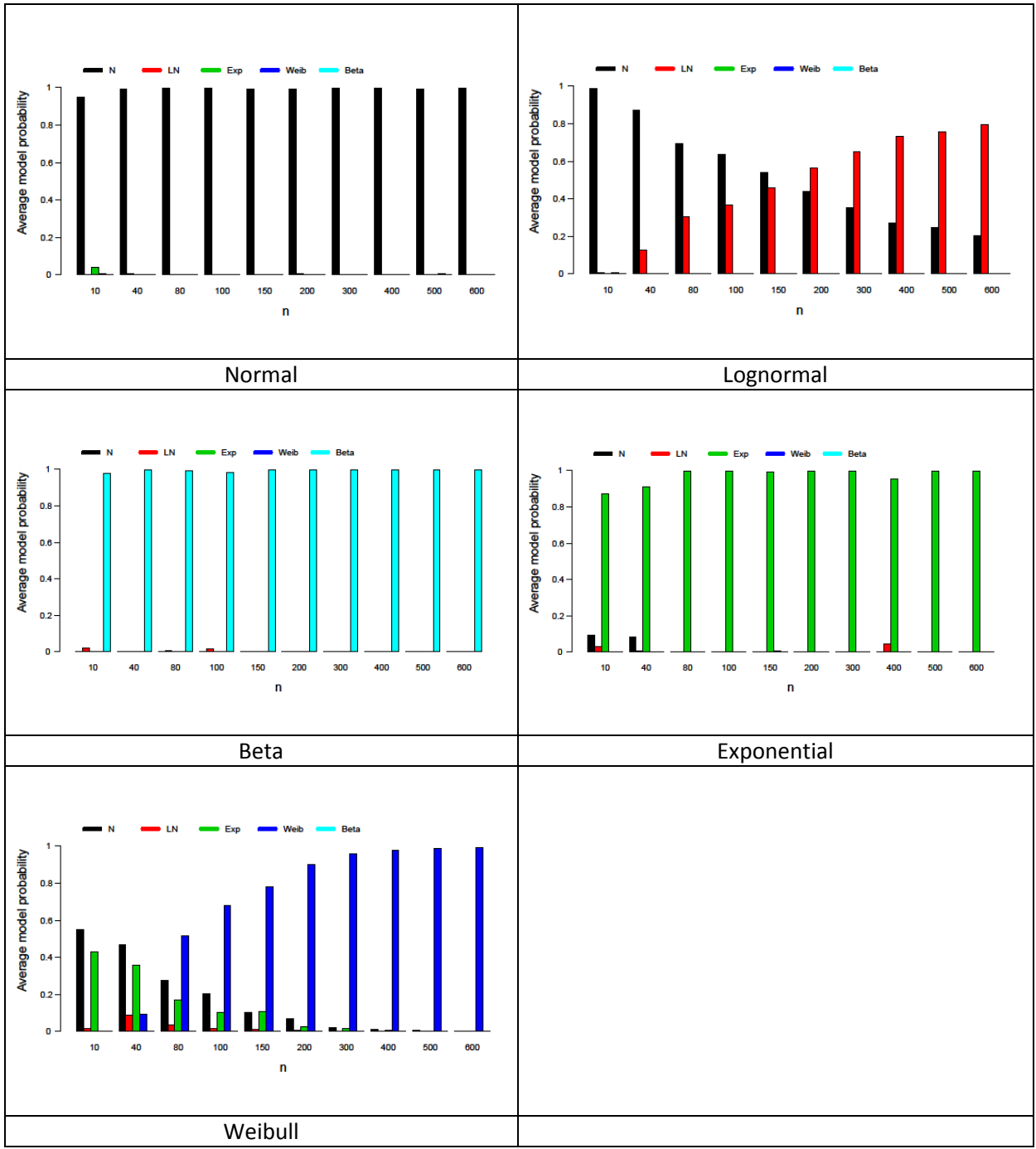

Normal

Lognormal

Beta

Exponential

Weibull

Figure 5B

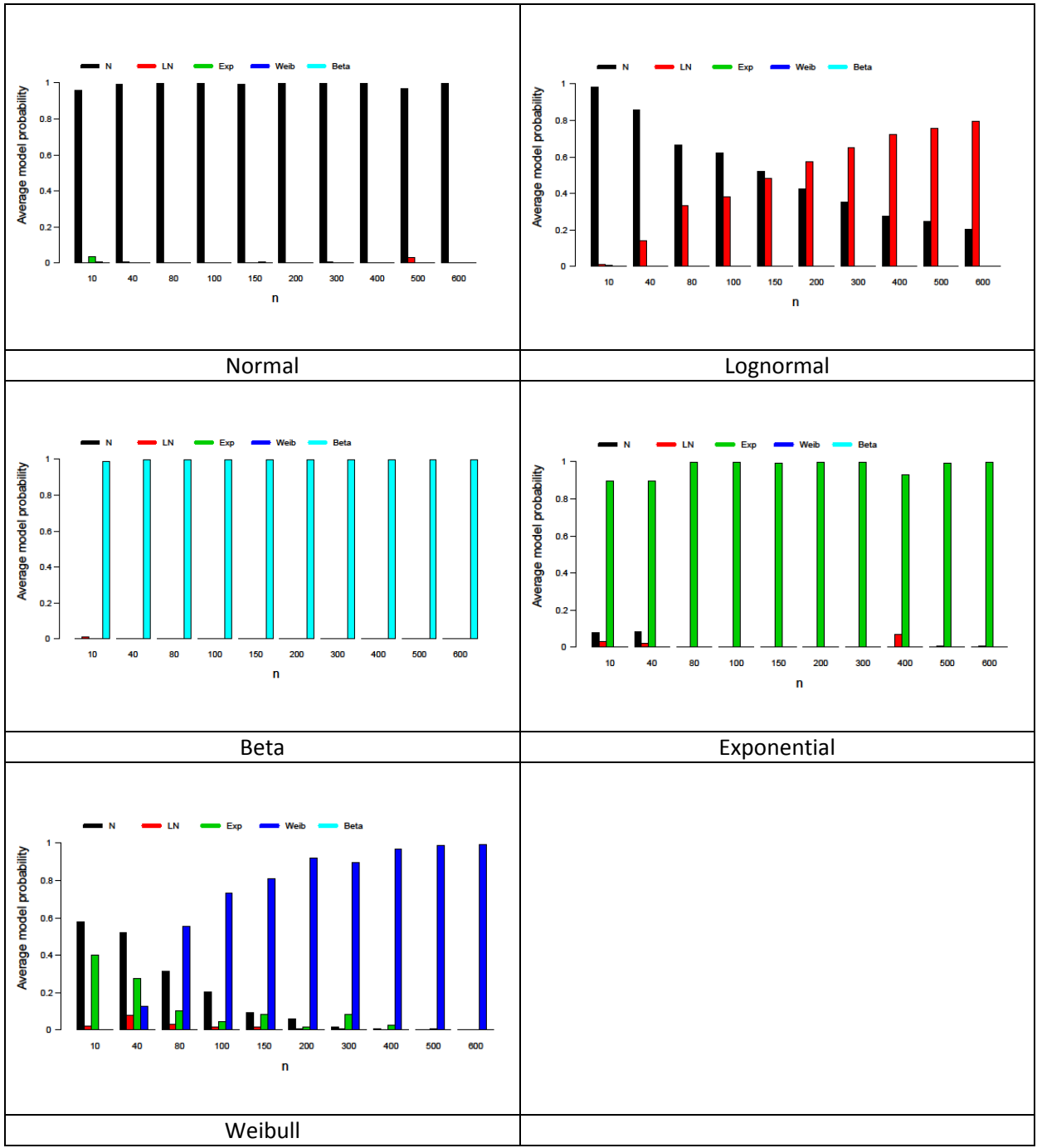

Figure 5C

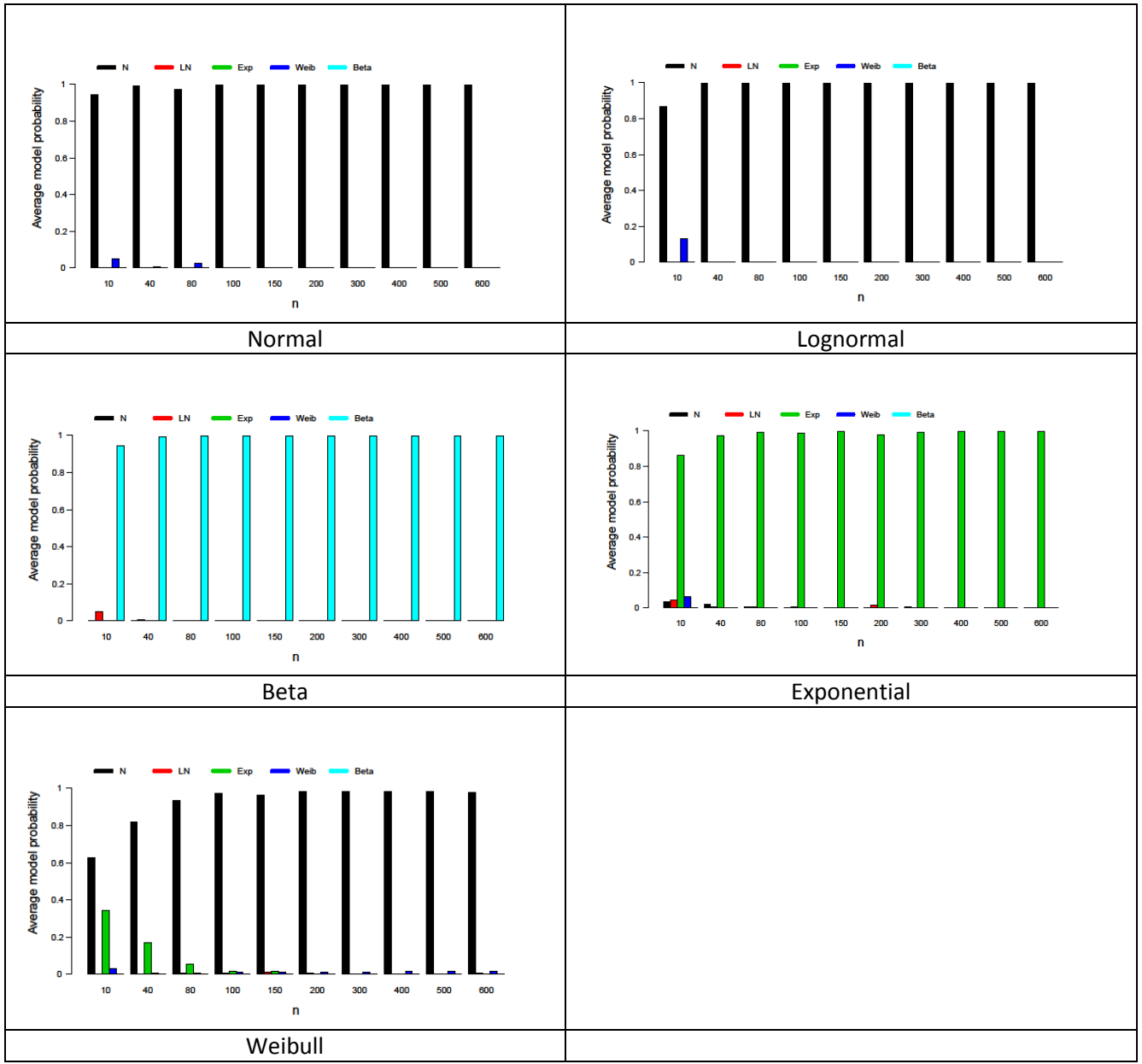

Figure 6

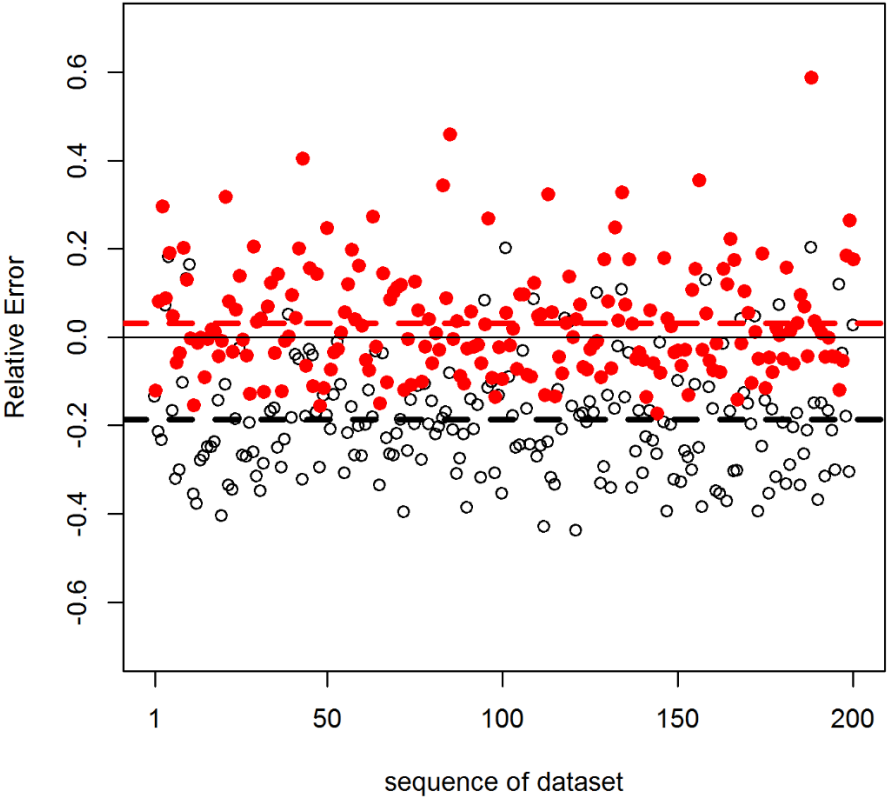

Table 1.  Procedures for implementation of ABC-SD and ABC-BMA methods

| ABC-SD | ABC-BMA |
|---|---|
| **Step 1:** Select a single distribution for underlying original data, f(θ). (See note below.) . <br><br> **Step 2:** θ* ~ p(θ); generate θ* from  uniform prior distribution(s). For example, if f is normal, uniform prior distributions for μ and σ need to be specified. <br><br> **Step 3:** D* ~ f(θ*); generate pseudo data. Compute summary statistics from pseudo data, S(D*)and compare with reported summary  statistics, S(D). <br><br> **Step 4:** If d(S(D*),S(D))< ε then ***θ\* is accepted***. <br><br>     Repeat steps 2-4 many times to get enough number of accepted θ* values (e.g., 20,000). <br><br> **Step 5:** Mean and standard deviation estimates are obtained using either 'plug-in method' or 'simulation'. <br><br> ------------------------ <br> **Note:** The single distribution for underlying distribution of original data is either an educated guess or it is determined via model selection using posterior probability criterion. | **Step 1:** Prepare a list of candidate distributions for underlying original data. <br><br> **Step 2:** Select one distribution $M_k$ from a set of K potential underlying distributions using multinomial ($\underline{\pi}$), where $\underline{\pi}=\{\pi_1,\ldots, \pi_K\}$[#] such as $\pi_k>0$ k=1,…,K and $\sum_k\pi_k=1$. Note: initial value of $\underline{\pi}=\{1/K,\ldots,1/K\}$. <br><br> **Step 3:** θ* ~ $p_{M_k}(\theta)$; generate θ* from corresponding prior  distribution of $M_k$. <br><br> **Step 4:** D* ~ $f_{M_k}(\theta^*)$; generate pseudo data. Compute summary statistics, S(D*) and compare with reported summary statistics, S (D). <br><br> **Step 5:** If d(S*, S)< ε then $M_k$ and estimates of mean and standard deviation are accepted. <br><br>     [#]Adjust π at every 1,000 iterations based on accepted frequency of each distribution. <br><br>     Repeat steps 2-5 many times to get enough number of accepted $M_k$ and estimates of mean and standard deviation (e.g., 50,000). <br><br> **Step 6:** Mean and standard deviation estimates are calculated as the weighted average of those accepted values (estimated means and standard deviations of pseudo-data). The weights are the proportions of accepted values of each candidate distribution, which are posterior model probabilities for each candidate distribution. |

Table 2. Sensitivity of prior distributions on estimation of mean and standard deviation.

| Prior setting | | Posterior model probability | | RE of mean | | | RE of SD | | |
| | | | | ABC-SD | | ABC-BMA | ABC-SD | | ABC-BMA |
| Beta: 2 shape parameters | Normal: σ parameter | Beta | Normal | Beta | Normal | -- | Beta | Normal | -- |
|---|---|---|---|---|---|---|---|---|---|
| U(0,40) | U(0,1) | 0.557 | 0.443 | -0.0021 | 0.0070 | 0.0019 | -0.0306 | 0.0423 | 0.0017 |
| U(0,40) | U(0,0.5) | 0.371 | 0.626 | -0.0025 | 0.0072 | 0.0036 | -0.0244 | 0.0433 | 0.0182 |
| U(0,20) | U(0,1) | 0.807 | 0.193 | -0.0020 | 0.0072 | -0.0080 | -0.0209 | 0.0463 | -0.0002 |
| U(0,20) | U(0,0.5) | 0.678 | 0.322 | -0.0021 | 0.0064 | 0.0007 | -0.0192 | 0.0448 | 0.0014 |

True sample estimates mean=0.6814 and SD=0.1247 in a simulated sample of 400 from Beta (9, 4).

RE:  relative error = *(estimated value − true value) / (true value)*.